\documentclass[10pt,journal,doublecolumn]{IEEEtran}
\usepackage{graphicx}
\usepackage{algorithm}
\usepackage{algorithmic}

\usepackage{booktabs}

\usepackage[colorlinks,
linkcolor=black,
anchorcolor=black,
citecolor=black,
urlcolor = blue
]{hyperref}

\usepackage{amssymb}
\usepackage{subfigure}
\usepackage{multicol}
\usepackage{multirow}
\usepackage{amsmath}
\usepackage{bm}
\usepackage{cite}
\usepackage{gensymb}
\usepackage{amsfonts}
\usepackage{mathrsfs}
\usepackage{amsmath}
\usepackage{color}
\usepackage{balance}
\usepackage{bm}
\usepackage{setspace}

\usepackage{stfloats}
\usepackage{float}
\usepackage{epstopdf}

\usepackage{array}
\newcommand{\PreserveBackslash}[1]{\let\temp=\\#1\let\\=\temp}
\newcolumntype{C}[1]{>{\PreserveBackslash\centering}p{#1}}
\usepackage[flushleft]{threeparttable}
\usepackage{stfloats}
\usepackage{mathrsfs}
\usepackage{threeparttable}
\usepackage{setspace}
\usepackage{hhline}
\usepackage{booktabs,tabularx}

\begin{document}

\bibliographystyle{IEEEtran}

\title{Optical IRS for Visible Light Communication: Modeling, Design, and Open Issues}

\author{Shiyuan Sun, Fang Yang, \emph{Senior Member, IEEE}, Weidong Mei, \emph{Member, IEEE},\\ Jian Song, \emph{Fellow, IEEE}, Zhu Han, \emph{Fellow, IEEE}, and Rui Zhang, \emph{Fellow, IEEE}

\begin{spacing}{0.3}
	\thanks{
		\scriptsize 
		S. Sun and F. Yang are with the Department of Electronic Engineering, Tsinghua University, Beijing 100084, P. R. China.
		\textit{(Corresponding author: Fang Yang)}

		W. Mei is the National Key Laboratory of Wireless Communications, University of Electronic Science and Technology of China, Chengdu 611731, China. 
	
		J. Song is with the Department of Electronic Engineering, Tsinghua University, Beijing 100084, P. R. China, and also with the Shenzhen International Graduate School, Tsinghua University, Shenzhen 518055, P. R. China. 
		
		Z. Han is with the Department of Electrical and Computer Engineering at the University of Houston, Houston, TX 77004 USA, and also with the Department of Computer Science and Engineering, Kyung Hee University, Seoul, South Korea, 446-701. 
		
		Rui Zhang is with School of Science and Engineering, Shenzhen Research
		Institute of Big Data, The Chinese University of Hong Kong, Shenzhen,
		Guangdong 518172, China, and also with the Department of Electrical and
		Computer Engineering, National University of Singapore, Singapore 117583.
	}
\end{spacing}
}
\newenvironment{thisnote}{\par\color{blue}}{\par}

\maketitle
\begin{abstract}
Optical intelligent reflecting surface (OIRS) offers a new and effective approach to resolving the line-of-sight blockage issue in visible light communication (VLC) by enabling redirection of light to bypass obstacles, thereby dramatically enhancing indoor VLC coverage and reliability.
This article provides a comprehensive overview of OIRS for VLC, including channel modeling, design techniques, and open issues. 
First, we present the characteristics of OIRS-reflected channels and introduce two practical models, namely, optics model and association model, which are then compared in terms of applicable conditions, configuration methods, and channel parameters.
Next, under the more practically appealing association model, we discuss the main design techniques for OIRS-aided VLC systems, including beam alignment, channel estimation, and OIRS reflection optimization.
Finally, open issues are identified to stimulate future research in this area.
\end{abstract}

\IEEEpeerreviewmaketitle
\begingroup
\allowdisplaybreaks

\section{Introduction}
\label{Sec:Intro}
Visible light communication (VLC) represents a cutting-edge advancement in wireless communication technology, by reusing the ubiquitous light-emitting diode (LED) lights that brighten our environments into conduits for high-speed data connectivity. 
This innovative approach leverages the vast unoccupied spectrum of visible light, which is anticipated to resolve the spectrum scarcity issues encountered in conventional radio frequency (RF) communications~\cite{chi2020visible}. 
In addition, the VLC can be realized by simply reusing our lighting infrastructure—a system that is already in place and powered, thus dramatically reducing the implementation complexity.
Despite the above benefits, VLC is susceptible to line-of-sight (LoS) blockage due to e.g., walls and furniture in the environment, which can significantly degrade the communication performance.

To circumvent the above limitations, optical intelligent reflecting surface (OIRS) has been proposed as a potential solution to the LoS blockage issue in VLC systems. 
Specifically, OIRS is a reconfigurable array that is capable of reflecting and redirecting light beams and can be implemented through mirror array, liquid crystal, or optical functional meta-material~\cite{9614037}.
By this means, the VLC signals can bypass the dense obstacles and reach their intended destinations even in the absence of a direct LoS path. 
By effectively ``bending'' light around corners and into shaded areas, OIRS can extend the effective signal coverage and achieve consistent and robust connectivity in VLC~\cite{jamali2021intelligent}.
Therefore, the integration of OIRS into VLC presents a compelling prospect by providing new degrees of flexibility and efficiency for VLC system design.

To date, two primary OIRS channel models have been widely adopted for VLC, namely the optics model and the association model~\cite{9968053}.
The former model characterizes the channel response in terms of the OIRS's actual parameters, which allows for a precise evaluation of the OIRS-reflected VLC channel gain but entails a high complexity for analysis.
The latter model, however, characterizes the channel gain in terms of the associations between the OIRS reflecting elements and the LEDs/photodetectors (PDs), which help ease the analysis. 
Based on the above two channel models, there have been many related works investigating the OIRS-assisted VLC for different purposes, such as channel modeling and analysis~\cite{ajam2023optical,9681888,abdelhady2020visible}, OIRS reflection optimization~\cite{sun2022mimo}, etc.
However, to the best of our knowledge, there is no existing work that thoroughly compares these two models for performance analysis of OIRS-assisted VLC.

Inspired by the above, this article first presents the characteristics of the OIRS-reflected VLC channel in different domains and unveil its unique properties as compared to the conventional RF IRSs, including angular selectivity, ``additive'' power, spatial coherence, and nanosecond delay.
Based on them, the optics and the association-based OIRS channel models are presented and compared in terms of their applicable conditions, configuration method, and channel state information (CSI) parameters.
Next, we proceed to discuss the main design techniques for OIRS-assisted VLC under the association-based OIRS channel model due to its appealing simplicity and accuracy.
Finally, open issues in the OIRS-assisted VLC are highlighted to stimulate future research.

\section{OIRS Channel Characteristics}
\label{Sec:Characteristics}
Due to the unique properties and constraints of the lightwave and optical devices, the OIRS-reflected channel exhibits many different characteristics from those of the RF IRS-reflected channel in various domains, as shown in Fig.~\ref{Fig:Channel} and discussed in the following.

\begin{figure*}[t]
	\centering
	\includegraphics[width=0.9\textwidth]{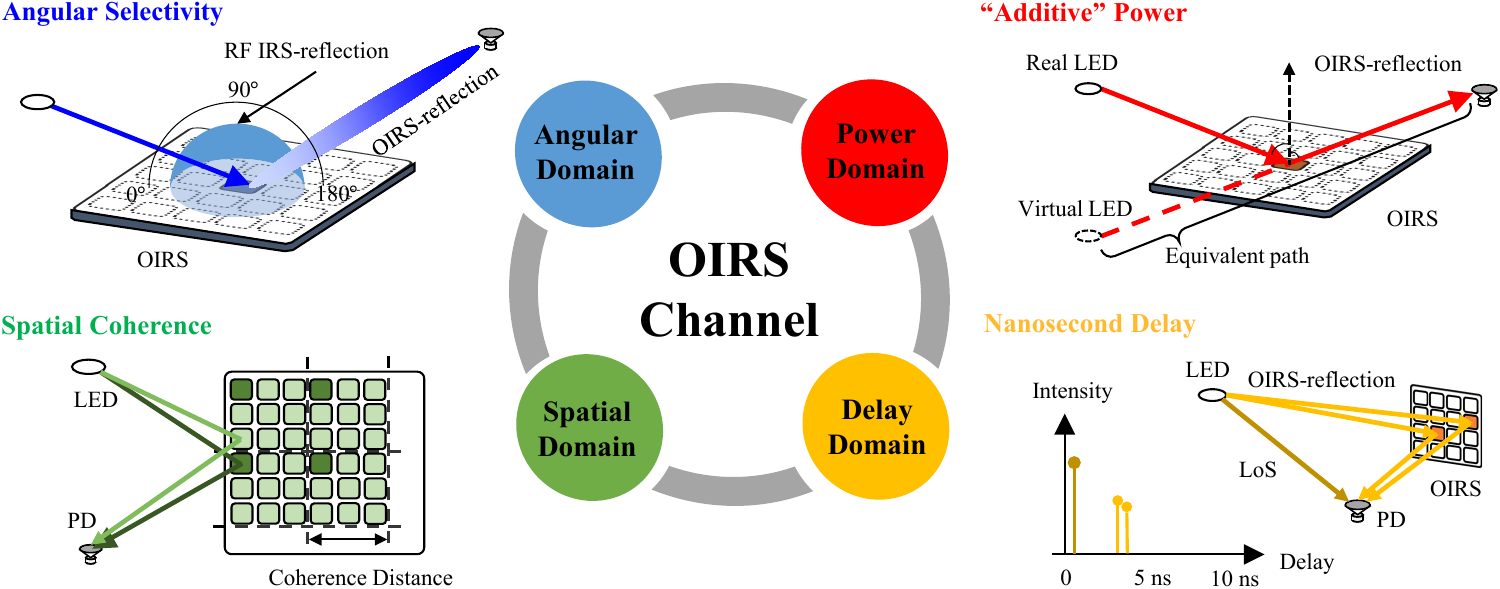}
	\caption{OIRS channel characteristics in the angular, power, spatial, and delay domains.}
	\label{Fig:Channel}
\end{figure*}

\subsection{Angular Domain}
\label{Subsec:Characteristics Angular}
According to computational electromagnetics, the reflection property of the impinging signal is dominated by the electrical size, which equals the ratio of the aperture of an IRS reflecting element to the signal wavelength~\cite{ozdogan2019intelligent}.
Conventional RF IRSs are comprised of numerous sub-wavelength elements (electrical size is smaller than $1$) and the spacing between two adjacent elements is typically half wavelength.
Under this condition, the energy distribution of the reflected electromagnetic wave tends to be isotropic in the angular domain, based on which each IRS element in the RF range can be approximated as a diffuse reflector.
Therefore, the beamforming of the RF IRS requires cooperatively tuning amplitudes and/or phase shifts of its reflecting elements.

However, as the electrical size becomes larger, the energy distribution changes from isotropic to anisotropic, making the diffuse reflector approximation for RF IRSs no longer valid.
Especially, the electrical size in VLC can be extremely large due to the nanoscale wavelength of the lightwave.
In this case, the energy of the reflected lightwave concentrates on a narrow mainlobe that follows the reflection law.
The energy distribution decays rapidly outside the mainlobe, such that the energy leakage to other directions can be ignored~\cite{jamali2021intelligent}.
This characteristic is termed as the \textit{angle-selective property} of the OIRS.
Accordingly, each OIRS reflecting element can be approximated as a specular reflector, and independent beams can be realized and controlled by all of the OIRS reflecting elements, which significantly differs from the RF IRS.

\subsection{Power Domain}
\label{Subsec:Characteristics Power}
Under the diffuse reflector approximation, the energy of the incident RF signal is reflected and dispersed uniformly in all directions.
Thus, the received energy at the IRS can experience significant attenuation over its surface, and the RF IRS-reflected channel can be characterized as a cascaded channel composed of two distinct sub-channels.
According to the physical optics, the overall gain of this cascaded channel is nearly the product of the gains of the two sub-channels, as well as the reflection coefficient of the IRS reflecting element, which is described as a ``multiplicative'' channel~\cite{wu2019towards}.

In contrast, the OIRS-reflected channel can be described as an ``additive'' channel.
According to the specular reflector approximation, the material of each element is considered homogenous and the element's aperture is sufficiently large relative to the light wavelength.
Consequently, geometric optics rather than physical optics is more applicable to analyze the OIRS-reflected channel, i.e., the reflected signal can be regarded as being directly emitted from a virtual LED, which is symmetrical to the actual LED across the plane of the OIRS reflecting element.

\begin{figure*}[t]
	\centering
	\includegraphics[width=0.9\textwidth]{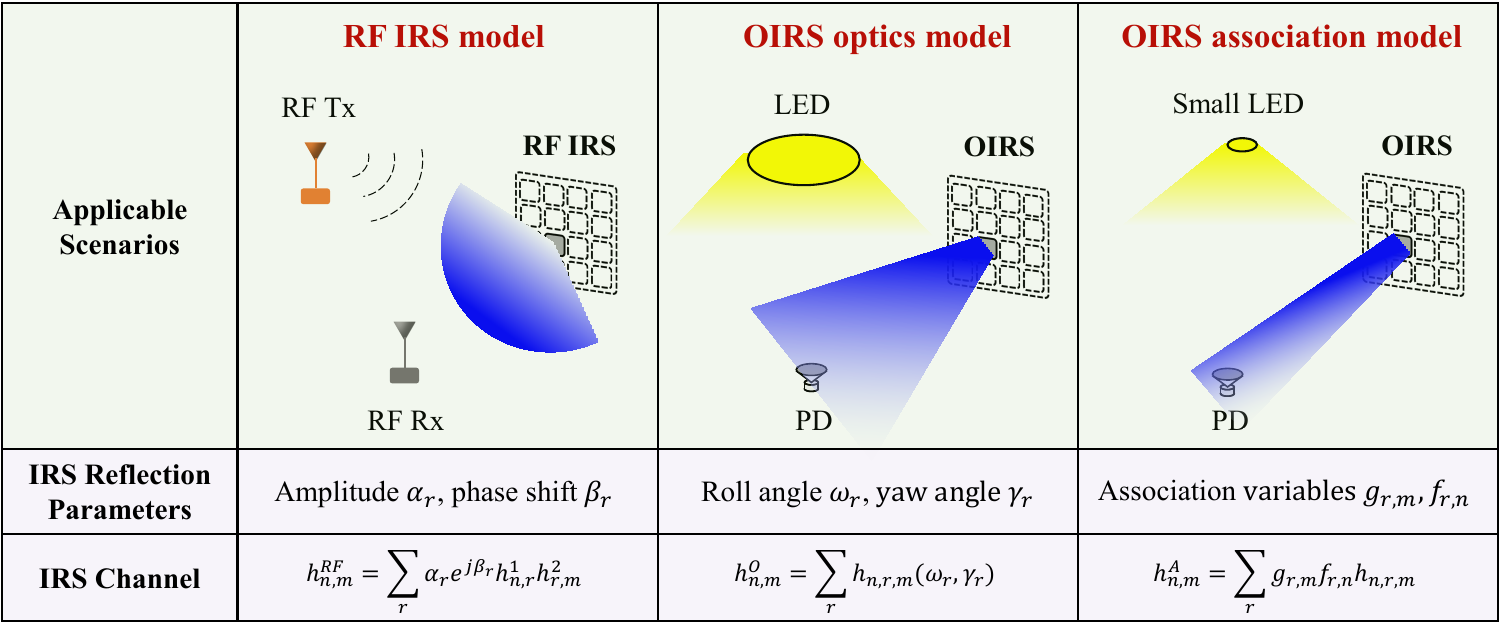}
	\caption{Comparisons among the RF IRS-reflected channel model, the OIRS optics model, and the OIRS association model.}
	\label{Fig:Models}
\end{figure*}

\subsection{Spatial Domain}
\label{Subsec:Characteristics Spatial}
In RF communications, both the channel gain and the transmitted signal are represented by complex numbers, and the channel gains can be decorrelated at adjacent RF IRS reflecting elements with certain (e.g., half-wavelength) spacing.
In contrast, the intensity modulation and direct detection (IM/DD) scheme is typically adopted in VLC, where only the amplitude of lightwave rather than its phase is utilized to modulate symbols.
This approach imposes a real-valued and non-negative constraint on the VLC channel gain and the transmitted signal.
For instance, the classic asymmetrically clipped optical-orthogonal frequency division multiplexing (ACO-OFDM) scheme meets this constraint by assigning zero symbols to the even subcarriers in the frequency domain, and then applying a clipping strategy in the time domain.

Due to the absence of the signal phase and the non-negative constraint, the impact of small-scale fading in VLC is less significant compared to that observed in RF communications since the amplitude fluctuation is slow for the VLC channel.
More specifically, the VLC channel gain, based on the Lambertian model, is contingent upon geometric parameters and the relative locations of the transceivers, including their areas, orientations, and the distance between them.
Generally, the channel gains associated with an LED and PD pair at adjacent OIRS reflecting elements are similar due to their proximity, which is so-called the spatial coherence of the OIRS-reflected channel.
In~\cite{sun2024estimation}, the authors have derived the OIRS coherence distance in closed form, which is defined as the maximum distance over which the growth rate of the reflected channel gain with respect to (w.r.t.) the shift on the OIRS stays below a certain threshold.
By exploiting this spatial coherence property, more efficient OIRS channel estimation can be achieved, as will be discussed later.

\subsection{Delay Domain}
\label{Subsec:Characteristics Delay}
For RF IRS-assisted systems, the delay spread for the LoS path, the IRS-reflected paths, and the diffusely reflected paths may result in frequency-selective channels.
However, in VLC, the diffusely reflected signals can be ignored due to the energy attenuation at the interface.
It is thus reasonable to only consider the LoS signal and the OIRS-reflected signal.
As such, the difference in propagation distance among multiple VLC paths is typically in the order of meters, which is approximately two orders of magnitude smaller than the scale encountered in RF communications (e.g., hundred meters) and leads to a delay in the order of nanoseconds~\cite{9681888}.
It follows that the OIRS-reflected channel can be generally considered as frequency-flat, as the PD receivers can hardly distinguish multiple paths.

\section{Association Model versus Optics Model}
\label{Sec:Model}
Based on the aforementioned channel characteristics, the IRS models for both RF and VLC systems are shown in Fig.~\ref{Fig:Models}.
It can be observed that the IRS-reflected channel gain in RF follows a ``multiplicative'' form, where $\alpha_r$, $\beta_r$, $h_{n,r}^1$, and $h_{r,m}^2$ denote the amplitude modification, the phase shift, the first sub-channel gain, and the second sub-channel gain at the $r$-th IRS reflecting element, respectively.
While the two OIRS channel models, namely, the optics model and the association model, are different from that in RF, as detailed below.

\subsection{Optics Model}
\label{Sec:Model single}
We consider a general system with a multi-PD receiver, a multi-LED transmitter, and an OIRS.
Assume that the OIRS is manufactured by an array of mechanically-driven mirrors, where the $r$th element can rotate itself along the yaw angle $\gamma_r$ and roll angle $\omega_r$ independently.
Considering the path from the $m$th LED to the $n$th PD and reflected by the $r$th OIRS element, the irradiance of an arbitrary point on the detection plane can be obtained following the physical optics~\cite{abdelhady2020visible}.
By integrating the received energy at the position occupied by the PD, the reflected channel gain by the $r$th OIRS element can be derived accordingly, i.e., $h_{n, r, m}(\omega_r, \gamma_r)$ in Fig. 2, which depends on the rotation angles of the OIRS.
As such, the overall OIRS-reflected channel gain $h_{n,m}^O$ is given by the superposition of $h_{n,r,m}(\omega_r,\gamma_r)$ over all OIRS reflecting elements. 
It is noted that this model precisely characterizes the relationship between the roll and yaw angles of the OIRS reflecting elements and the reflected channel gain. 
Particularly, an OIRS reflecting element can make contributions to multiple PDs and LEDs.

\subsection{Association Model}
\label{Sec:Model multiple}
Unlike the precise channel modeling as in the optics model, the OIRS-reflected channel can be simplified given a small source.
Specifically, assuming that the size of the LED is much smaller than the propagation distance (which is usually the case in practice), the reflected energy will become concentrated within a narrow beam due to the angle-selective property discussed in Section~\ref{Subsec:Characteristics Angular}.
As shown in Fig.~\ref{Fig:Point}, the energy distribution on the detection plane forms four peaks, each corresponding to a different OIRS element.
It follows that a single OIRS reflecting element only affects one pair of LED and PD without causing inter-element interferences, i.e., for the $r$th OIRS reflecting element, there is at most one non-zero entry in the sets of $\{h_{n, r, m}\}_{n}$ and $\{h_{n, r, m}\}_{m}$.

Based on this observation, the OIRS-reflected channel can be modeled in terms of the associations between the OIRS reflecting elements and LEDs/PDs.
To depict the associations, we define binary parameters $g_{r,m}$ and $f_{r,n}$, which indicate the associations of the OIRS reflecting elements with the LEDs and with the PDs, respectively~\cite{sun2022mimo}.
Due to the angle-selective property, each OIRS reflecting element is associated with at most one LED and one PD. 
As such, the summation of $g_{r,m}$ w.r.t. $m$ and that of $f_{r,n}$ w.r.t. $n$ should be less than or equal to $1$.
As such, the channel gain of the association model can be expressed as $h^A_{n, m}(g_{r,m}, f_{r,n})$, as shown in Fig.~\ref{Fig:Point}.
Next, given the association results, the rotation angle of each OIRS reflecting element can be tuned to reeceive and reflect the incident beams based on its associated LED and PD, respectively.
As compared to the optics model, this model avoids capturing the sophisticated optical relationships and yields a concise expression for the OIRS-reflected channel.

\begin{figure}[t]
	\centering
	\includegraphics[width=0.45\textwidth]{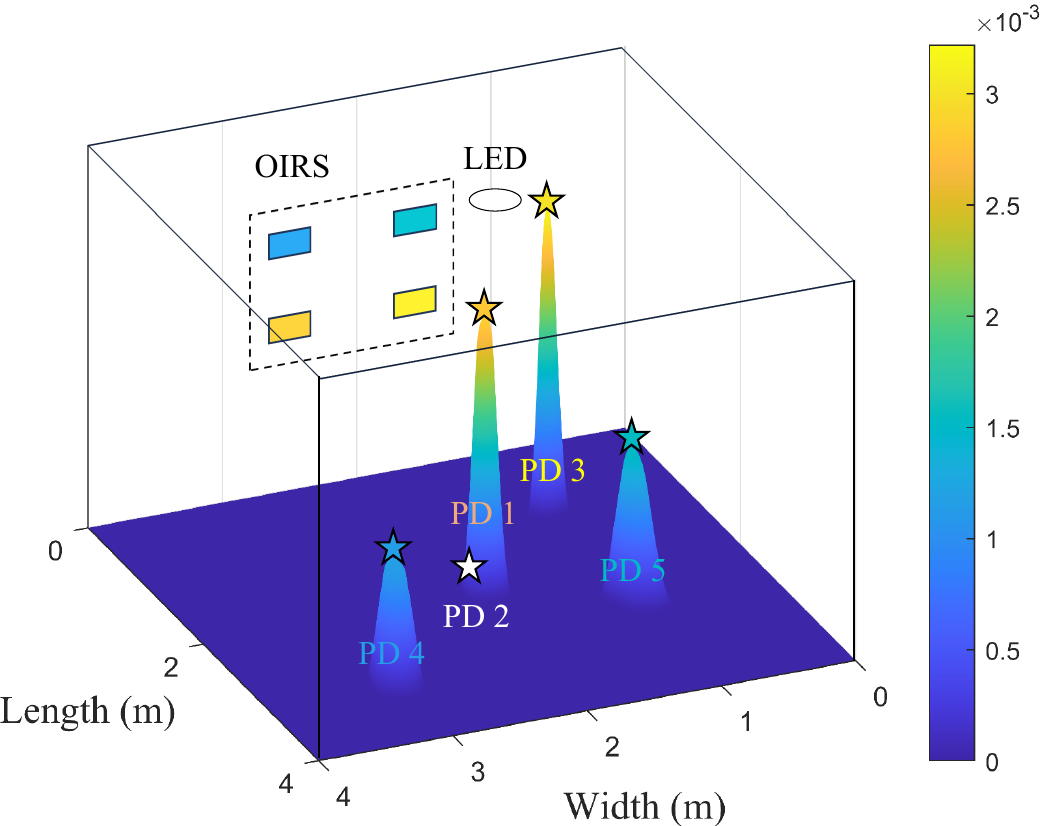}
	\caption{Energy distribution of the OIRS-reflected signal.}
	\label{Fig:Point}
\end{figure}

\subsection{Comparison Between the Two OIRS Models}
\label{Sec:Comparisons}
\textbf{Applicability:}
For the optics model, the OIRS-reflected channel gain is expressed in terms of the actual OIRS reflection parameters, e.g., the rotation angle and the voltage under the mirror-array and meta-material implementations, respectively.
Hence, this model is general and can be applied to any VLC systems with arbitrary sizes of the LED, PD, and OIRS.
Moreover, the OIRS reflection optimization under this model can be conducted w.r.t. the actual OIRS reflection parameters; thus, the optimized reflection can be directly adopted for OIRS configuration.
However, the optics model is nonlinear and integral, making it difficult to solve the related reflection optimization problem, especially in the multi-user and/or multi-antenna system.
Additionally, the optics model relies on the accuracy of the Lambertian radiation pattern, which can be imprecise in practice.

The OIRS association model can be considered as a simplified form of the optics model, under the prerequisite that the LED size should be considerably smaller than the propagation distance.
Particularly, it avoids the complicated optical formulation and renders a linear expression of the OIRS-reflected channel gain, thus greatly simplifying the OIRS reflection optimization.
Moreover, the OIRS association model is not derived from the Lambertian-based model, which thus avoids the inaccuracy issue with the optics model.

A numerical comparison between the the OIRS-reflected channel gains under the two models is depicted in Fig.~\ref{Fig:Simulation}.
We consider an OIRS reflecting element is tuned to form a reflected path from a single LED to one PD, and there is another PD with a spacing of 0.2 m from the first PD.
It is observed that the channel gains under the association model and the optical model, i.e., $h_{1,1}^A$ and $h_{1,1}^O$, are comparable. 
Nonetheless, $h_{1,1}^A$ is observed to be slightly smaller than $h_{1,1}^O$, which is mainly due to the imperfect beam alignment between the OIRS reflecting elements and the transceiver antenna pairs with finite rotational angle resolution. 
For the second PD, its reflected channel gain under the association model, i.e., $h_{2,1}^A$, is observed to be $0$ due to the constraint on $g_{r,m}$, i.e., each OIRS reflecting element can be associated with at most one PD. 
It is observed that as compared to the actual channel gain under the optics model (i.e., $h_{2,1}^O$), the modeling error is negligible when the radius of the LED is smaller than $0.15$ m.

\begin{figure}[t]
	\centering
	\includegraphics[width=0.45\textwidth]{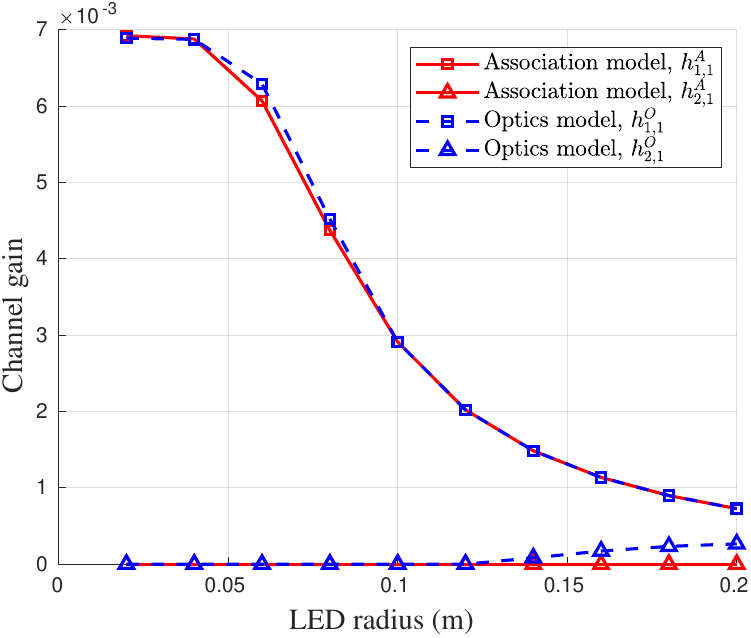}
	\caption{OIRS-reflected channel gains under the optics model versus the association model.}
	\label{Fig:Simulation}
\end{figure}

\textbf{OIRS Configuration:}
Under the optics model, the OIRS reflection parameters are the yaw angle and roll angle, which is directly related to the actual configuration of OIRS.
While the OIRS reflection parameters of the association model are the binary association variables between the OIRS reflecting elements and LEDs/PDs.
As such, the actual OIRS configuration in the association model needs to be determined by applying the beam alignment techniques for each OIRS reflecting element and its associated LEDs and PDs, which is to be specified in Section~\ref{Subsec:Technologies Management}.

\textbf{CSI Parameters:}
In the optics model, the channel gain of the OIRS-reflected path is derived based on the Lambertian radiation pattern and can be expressed as~\cite[Eq. (51)]{abdelhady2020visible}.
Therefore, the CSI is represented by geometric parameters, such as the propagation distances before and after the reflection, the angles of incidence and reflection at the LED/PD/OIRS.
Note that these CSI parameters can be intricately coupled with the OIRS reflection parameters in an integral form, which thus poses a significant challenge for CSI acquisition.
In the association model, however, the OIRS-reflected channel is expressed in a linear form, such that the CSI parameters can be represented by $h_{n, r, m}$ directly and efficiently acquired based on the pre-designed OIRS reflection pattern and pilots.
The comparisons between the two OIRS-reflected channel models are summarized in Table~\ref{tab:tab1}.

\newcommand{\tabincell}[2]{
	\begin{tabular}{@{}#1@{}}#2\end{tabular}
}
\renewcommand\arraystretch{1.1}
\begin{table*}[t]
	\centering
	\caption{Comparison among different IRS-reflected channel models}
	\renewcommand{\tabularxcolumn}[1]{m{#1}}
	\begin{tabularx}{\textwidth}{|m{2.5cm}|X|X|X|}
		\hline
		\multicolumn{1}{|c|}{\textbf{\small Properties}}  & \multicolumn{1}{c|}{\textbf{\small RF IRS model}} & \multicolumn{1}{c|}{\textbf{\small OIRS optics model}} & \multicolumn{1}{c|}{\textbf{\small OIRS association model}} \\ 
		
		\hline
		\textbf{Angular domain} & Isotropic (diffuse reflector) & \multicolumn{2}{c|}{Anisotropic (specular reflector)} \\
		\cline{1-4}
		\textbf{Power domain} & ``Multiplicative'' & \multicolumn{2}{c|}{``Additive''} \\ 
		\cline{1-4}
		\textbf{Spatial domain} & Less coherence & \multicolumn{2}{c|}{Coherence} \\ 
		\cline{1-4}
		\textbf{Delay domain} & Order of microseconds & \multicolumn{2}{c|}{Order of nanoseconds} \\ 
		
		\hline
		\textbf{Applicability} & RF range and sub-wavelength element & Optical range & Optical range and small source \\ 
		\cline{1-4}
		\textbf{Configuration} & Set amplitude and phase shifts & Set yaw and roll angles & \tabincell{l}{Set association variables and \\  conduct beam alignment} \\ 
		\cline{1-4}
		\textbf{CSI parameters} & First and second sub-channel gains & \tabincell{l}{Geometric parameters and \\ locations of the LEDs/PDs} & \tabincell{l}{Optical channel gains of all \\ LED-OIRS-PD pairs} \\
		\hline
	\end{tabularx}
	\label{tab:tab1}
\end{table*}

\section{Key Design Techniques for OIRS-assisted VLC\\ under the Association Model}
\label{Sec:Technologies}
Considering the practically appealing balance between the modeling accuracy and simplicity of the association model, we overview the key techniques for designing the OIRS-assisted VLC in this section under this model.
An OIRS-assisted VLC protocol, including beam alignment, channel estimation, and OIRS reflection optimization, is shown in Fig.~\ref{Fig:Protocol}, where the communication process can be divided into the CSI acquisition stage and the channel use stage, which are elaborated in detail in the next.

\subsection{Beam Alignment}
\label{Subsec:Technologies Management}
First, given the association results, the rotation angles of each OIRS reflecting element should be tuned to align with its associated LED/PD pair to configure the OIRS.
To this end, a set of rotational angles (i.e., roll and yaw angles) can be generated to form a angle codebook by sampling the angular space of each OIRS element.
As such, the OIRS beam alignment can be realized by searching over the angle codewords.

However, the performance of the OIRS codebook critically depends on the sampling strategy for the roll angle and yaw angle.
The most straightforward strategy is to uniformly and separately sample the rotational angles, and the resulting angle codebook can be regarded as the Cartesian product of two constituent codebooks corresponding to the roll angle and yaw angle, respectively.
Such an OIRS codebook based on uniform sampling can achieve a high alignment resolution if the sampling spacing is sufficiently small, whereas the number of angle codewords may become extremely large.
To reduce the searching complexity, a non-uniform sampling strategy is proposed in~\cite{sun2024estimation} based on the geometric optics.
The rationale is that the distribution of uniform codewords is denser in proximity to the OIRS reflecting element and becomes sparser with increasing distance.
Therefore, both of the two angles can be samplied in a non-uniform fashion to realize uniform distribution of the codewords on the detection plane, as depicted in the leftmost figure in Fig.~\ref{Fig:Protocol}.

\subsection{Channel Estimation}
\label{Subsec:Technologies estimation}
CSI acquisition is crucial for optimizing the OIRS reflection, yet it remains an unresolved challenge under the optical model.
This is mainly due to the intricate coupling of the CSI parameters and OIRS reflection parameters in the complex channel expression. 
To circumvent this difficulty, a data-driven algorithm has been proposed for OIRS channel estimation under this model~\cite{10468699}.
While in the association model, for any given LED-OIRS-PD associations and beam alignment, the received signal at each PD shares a similar structure to the conventional RF model. 
As such, various classical channel estimation techniques such as minimum mean square error (MSE) estimator can be utilized for OIRS channel estimation.

Nonetheless, one critical challenge lies in the fact that the number of CSI parameters in the association model is practically large (considerably larger compared to the RF IRS) as it involves the reflected channel gains for any LED-OIRS-PD pair.
To tackle this problem, a space-time sampling algorithm has been proposed in~\cite{sun2024estimation} to reduce the number of estimated parameters by leveraging the coherence characteristics of the OIRS channels in the time and spatial domains. 
However, there is still room for more in-depth research into more efficient techniques.

\begin{figure*}[t]
	\centering
	\includegraphics[width=0.9\textwidth]{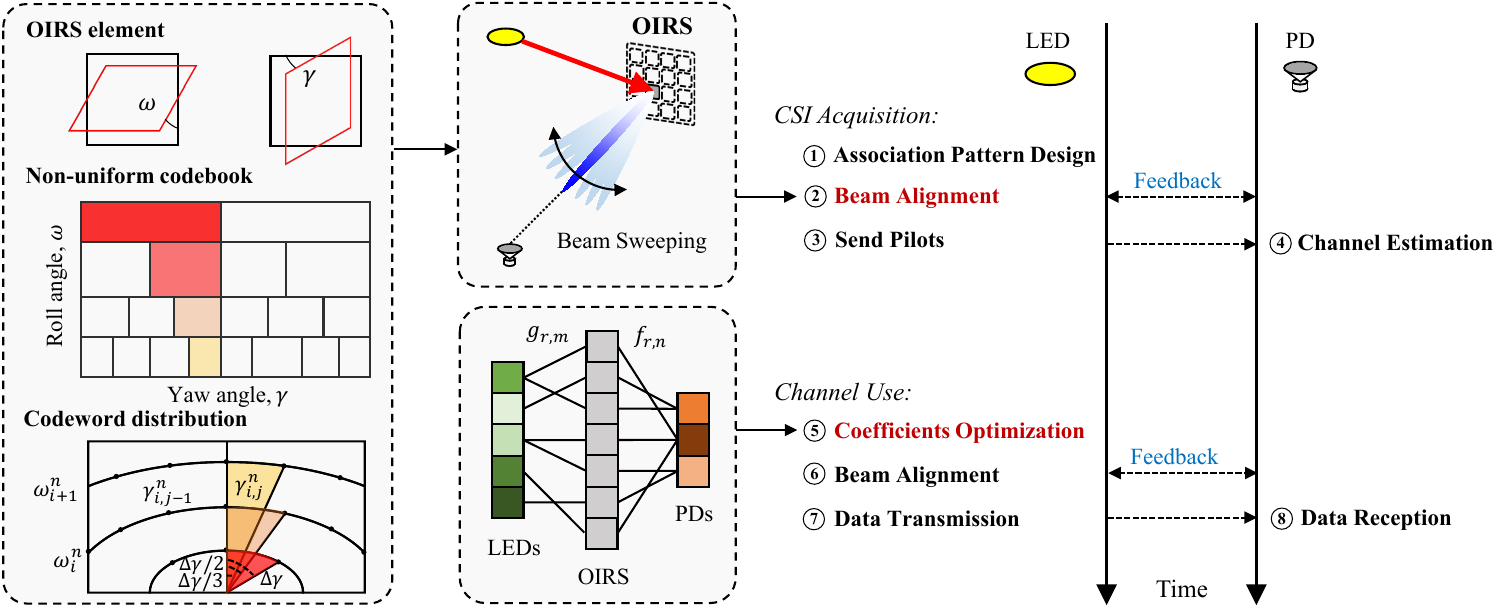}
	\caption{Beam alignment, codebook design, and reflection optimization under the association-based OIRS channel model.}
	\label{Fig:Protocol}
\end{figure*}

\subsection{Multi-Antenna and Multi-User Systems}
\label{Subsec:Technologies Optimization}
The OIRS association model can be easily applied to more general cases with multiple antennas and/or users.
For example, in the multiple-input single-output (MISO) system, the OIRS association variable $f_{r,n}$ must be one. 
Hence, it suffices to optimize the associations between the OIRS reflecting elements and the LEDs.
Similarly, the results can be applied to the single-input multiple-output (SIMO) setup due to its duality with the MISO setup. 
Moreover, in the multiple-input multiple-output (MIMO) system, the effective MIMO channel can be regarded as being composed of the baseband precoding module at the transmitter, the OIRS-reflected MIMO channel, and the detection module at the receiver.
Based on this structure, the MSE minimization problem has been investigated by optimizing the above three components in an alternate manner~\cite{sun2022mimo}.
Finally, for the multi-user (MU)-MIMO system, the associations between the OIRS reflecting elements and the PDs should cater to different users, which is thus more difficult to be solved as compared to the single-user case.

\subsection{Reflection Optimization}
\label{Subsec:Technologies optimization}
The OIRS reflection optimization in the association model allows for the use of more efficient optimization algorithms, as compared to that in the optics model thanks to the simpler channel expressions~\cite{10436776}. 
However, owing to the binary association variables involved in this model, the OIRS reflection design is a discrete optimization problem.

A variety of optimization algorithms can be utilized to solve such a discrete problem.
The most straightforward method is the brute-force optimization algorithm, which ensures global optimality but entails prohibitively high complexity especially if the number of OIRS reflecting elements is large.
An alternative method is to transform the discrete problem into a continuous problem by relaxing the discrete OIRS association variables, for which classical optimization algorithms such as the gradient descent algorithm, successive convex approximation, and alternating optimization can be applied.
Moreover, the discrete association variables can be retrieved from the optimized continuous variables based on the minimum projection error criterion.
In addition to the above two methods, it has been shown in~\cite{sun2022mimo} that the OIRS association variables w.r.t. the LEDs and PDs can be replaced equivalently by a composite variable based on the Hadamard product.
By leveraging this variable substitution, the OIRS reflection optimization problem can be greatly simplified, and the problem can be solved by exploiting one of the above two algorithms.

\section{Open Issues}
\label{Subsec:Technologies Open}
The investigation into OIRS under the association model is still in an early stage, and a number of remaining issues need to be resolved, as listed below.

\textbf{Efficient Beam Alignment:}
Beam alignment is a basis of the OIRS association model and needs to be maintained during the whole communication process.
To this end, the design of the OIRS angle codebook may be improved from the current single-layer codebook (uniform or non-uniform) to a multi-layer structure, which helps reduce the searching time.
Moreover, the efficiency of the beam sweeping strategy may be refined based on the coarse location information since the OIRS-reflected channel follows the geometric optics.
Particularly, beam tracking can be performed to maintain the beam alignment that is already established.

\textbf{Uplink Optimization:}
Existing works mostly focus on the downlink of the OIRS-reflected VLC channel.
However, as a bi-directional communication technology, light fidelity (LiFi) requires both downlink and uplink optical wireless communications in one system.
Particularly, the uplink OIRS-assisted communication differs from its downlink counterpart in several aspects, e.g., individual power constraint and the adopted invisible frequency band.
The uplink channel model can be established according to the principle of optics reversibility, and the OIRS reflection optimizations for uplink requires further investigation.

\textbf{Integration with Other Technologies:}
The integration of OIRSs with other technologies is promising due to the additional design degree-of-freedom, e.g., integrating non-orthogonal multiple access (NOMA) to improve the data rate and integrating visible light positioning to enhance the positioning performance.
In addition, OIRS can be utilized to create artificial interference towards eavesdroppers without compromising the performance of legitimate users, thereby enhancing the physical layer security (PLS) of VLC.
Moreover, OIRS is promising in improving the performance of simultaneous wireless information and power transfer (SWIPT) considering that the received energy can be increased due to the IM/DD scheme.
Efficient integration of OIRS and other technologies needs to be investigated in the future.

\textbf{Machine Learning (ML)-based OIRS Reflection Optimization:}
Considering that the OIRS optimization problem is a combinatorial programming problem and NP-hard, it is generally difficult to be optimally solved.
To overcome this challenge, the ML-based algorithms, a data-driven tool that can optimize the OIRS parameters and communication resources based on the pre-generated samples, turn out to be a viable solution.
There have been several related works utilizing reinforcement learning to optimize the OIRS reflection parameters, including autoencoder design and deep reinforcement learning (DRL)-based design~\cite{10468699,al2023deep}, etc.
It would be promising to develop more efficient ML-based optimization algorithms for the OIRS-assisted communication system.

\textbf{Prototype Manufacturing:}
Compared to theoretical analysis, the prototyping of the OIRS needs to be accelerated for both optical functional material implementation and mirror-array implementation.
The meta-lens with electrically stretchable artificial muscles and the liquid-crystal-based RIS with electronically adjustable refractive index are two typical instances of optical functional material, which can steer the incident lightwave dynamically for enhancing the signal detection capabilities of PD~\cite{ndjiongue2021toward}.
On the other hand, an OIRS prototype based on the mirror array employs the reflection law in optics, and each element performs as a reflector that can rotate in two independent angles.
More advanced OIRS prototypes in terms of the tunning principle, tunning accuracy, response delay, and manufacturing cost should be developed.

\section{Conclusions}
\label{Sec:Conclude}
This article investigated the OIRS-assisted VLC from the aspect of channel modeling.
Specifically, based on the channel characteristics in terms of angular/power/spatial/delay domains, two primary OIRS-reflected channel models, namely the association-based model and the optics-based model, were introduced and thoroughly compared in terms of applicability, configuration methods, and CSI parameters.
Next, we discussed the main techniques for designing the OIRS-assisted VLC under the more practically appealing OIRS association model and pointed out some promising directions for future OIRS research. 
With the integration of OIRS, VLC is anticipated to become a more reliable and cost-effective technology for realizing indoor optical wireless communications in the next generation of wireless networks.

\bibliographystyle{abbrv}
\bibliography{IEEEabrv,reference}

\end{document}